\documentclass[twocolumn,showpacs,superscriptaddress,preprintnumbers,amsmath,amssymb]{revtex4}
\usepackage{graphicx}% Include figure files
\usepackage{dcolumn}% Align table columns on decimal point
\usepackage{bm}% bold math
\usepackage{epsfig}
\begin{document}

\title{Relaxation optimized transfer of spin order in Ising spin chains}

\author{Dionisis Stefanatos}
\email{stefanat@fas.harvard.edu} \affiliation{Division of
Engineering and Applied Sciences, Harvard University, Cambridge,
Massachusetts 02138, USA}
\author{Steffen J. Glaser}
\affiliation{Department of Chemistry, Technische Universit\"{a}t
M\"{u}nchen, 85747 Garching, Germany}
\author{Navin Khaneja}
\affiliation{Division of Engineering and Applied Sciences, Harvard
University, Cambridge, Massachusetts 02138, USA}

\date{\today}

\begin{abstract}
In this manuscript, we present relaxation optimized methods for
transfer of bilinear spin correlations along Ising spin chains.
These relaxation optimized methods can be used as a building block
for transfer of polarization between distant spins on a spin
chain. Compared to standard techniques, significant reduction in
relaxation losses is achieved by these optimized methods when
transverse relaxation rates are much larger than the longitudinal
relaxation rates and comparable to couplings between spins. We
derive an upper bound on the efficiency of transfer of spin order
along a chain of spins in the presence of relaxation and show that
this bound can be approached by relaxation optimized pulse
sequences presented in the paper.
\end{abstract}

\pacs{03.67.-a, 03.65.Yz, 82.56.Jn, 82.56.Fk}
\maketitle

\section{Introduction}
Relaxation (dissipation and decoherence) is a characteristic
feature of open quantum systems. In practice, relaxation results
in loss of signal and information and ultimately limits the range
of applications. Recent work in optimal control of spin dynamics
in the presence of relaxation has shown that these losses can be
significantly reduced by exploiting the structure of relaxation
\cite{Khaneja1, Khaneja2, BBCROP, Stefanatos}. This has resulted
in significant improvement in sensitivity of many well established
experiments in high resolution nuclear magnetic resonance (NMR)
spectroscopy. In particular, by use of optimal control methods,
analytical bounds have been achieved on the maximum polarization
or coherence that can be transferred between coupled spins in the
presence of very general decoherence mechanisms. In this paper, we
look at the more general problem of transfer of coherence or
polarization between distant spins on an Ising spin chain in the
presence of relaxation. This problem is ubiquitous in
multi-dimensional NMR spectroscopy, where polarization is
transferred between distant spins on a chain of coupled spins.
Spin (or pseudo spin) chains also appear in many proposed quantum
information processing architectures \cite{kane, yama}.

\begin{figure}[t]
\centering
\includegraphics[scale=0.5,clip=true,viewport=1.5cm 4cm 18cm 7.7cm]{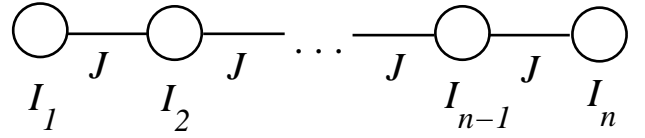}
\caption{\label{fig:chain}The system that we study in this paper
is a linear chain of $n$ weakly interacting spins $1/2$ with Ising
coupling between next neighbors. The coupling constant is the same
for all pairs of connected spins.}
\end{figure}

The system that we study in this paper is a linear chain of $n$
weakly interacting spins $1/2$ placed in a static external
magnetic field in the $z$ direction (NMR experimental setup), with
Ising type couplings of equal strength between nearest neighbors,
see Fig. \ref{fig:chain}. The free evolution Hamiltonian of the
system has the form
$$H_0 = \sum_{i=1}^n\omega_i I_{iz} + 2 \pi
J\sum_{i=1}^{n-1}I_{iz}I_{(i+1)z},$$ where $\omega_i$ is the Larmor
frequency of spin $i$ and $J$ is the strength of the coupling between the spins.
In a suitably chosen (multiple) rotating frame, which rotates with each
spin at its resonance (Larmor) frequency, the free evolution
Hamiltonian simplifies to
\begin{equation}
\label{Hamiltonian} H_{c}= 2 \pi
J\sum_{i=1}^{n-1}I_{iz}I_{(i+1)z}\;.
\end{equation}

Motivated by NMR spectroscopy of large molecules in solution, we
assume that the relaxation rates of the longitudinal operators
with components only in the $z$ direction, like $I_{iz}$,
$2I_{iz}I_{(i+1)z}$ is negligible compared with relaxation rates
for transverse operators like $I_{ix}$ , $2I_{iz}I_{(i+1)y}$
\cite{NMR}. The transverse relaxation is modeled by the
Lindbladian with the general form \cite{NMR}
$$ L(\rho) =\sum_i \pi a_i [I_{iz}[I_{iz}, \rho]] + \sum_{i,j}\pi b_{ij}
[2I_{iz}I_{jz}[2I_{iz}I_{jz}, \rho]]\;.$$ In liquid state NMR
spectroscopy, the two terms of the Lindblad operator model the
relaxation mechanism caused by chemical shift anisotropy and
dipole-dipole interaction respectively \cite{NMR}. Here we neglect
any interference effects between these two relaxation mechanisms
\cite{Goldman1}. Relaxation rates $a_i,b_{ij}$ depend on various
physical parameters, such as the gyromagnetic ratios of the spins,
the internuclear distance, the correlation time of molecular
tumbling etc. We define the net transverse relaxation rate for
spin $i$ as  \cite{Khaneja1} $k_a^{i} = a_i + \sum_j b_{ij}$.
Without loss of generality in the subsequent analysis, we assume
that $k_a^i$ are equal and we denote this common transverse
relaxation rate by $k$.

The time evolution (in the rotating frame) of the spin system
density matrix $\rho$ is given by the master equation
\begin{equation}\label{eq:master}
\dot{\rho}=-i[H,\rho]+L(\rho)\;,
\end{equation}
where $H=H_c+H_{rf}$ and $H_{rf}$ is the control Hamiltonian. In
the NMR context, the available controls are the components of the
transverse radio-frequency (RF) magnetic field. It is assumed that
the resonance frequencies of the spins are well separated, so that
each spin can be selectively excited (addressed) by an appropriate
choice of the components of the RF field at its resonance
frequency.

Consider now the problem of optimizing the polarization transfer
\begin{equation}
\label{transfer} I_{1z}\rightarrow I_{nz}
\end{equation}
along the linear spin chain shown in Fig. \ref{fig:chain}, in the
presence of the relaxation mechanisms mentioned above. This
problem can be stated as follows: Find the optimal transverse RF
magnetic field in the control Hamiltonian $H_{rf}$ such that
starting from $\rho(0)=I_{1z}$ and evolving under Eq.
(\ref{eq:master}) the target expectation value $\langle I_{nz}
\rangle=\mbox{tr}\{\rho(t)I_{nz}\}$ is maximized.

\begin{figure}[t]
\centering
\includegraphics[scale=0.5,]{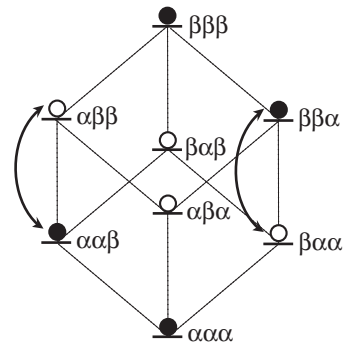}
\caption{\label{fig:zzlevels}Energy level diagram for the spin
order transfer $2I_{1z}I_{2z} \rightarrow 2I_{2z}I_{3z}$. The dark
circles represent population excess. State $\alpha$ corresponds to
spin up and state $\beta$ to spin down.}
\end{figure}

To fix ideas, we analyze the case when $n=3$
\begin{equation}
\label{transferpolarization}
I_{1z}\rightarrow I_{3z}\;.
\end{equation}
This transfer is achieved conventionally using INEPT like pulse
sequences \cite{inept1, inept2, concat}. Under the conventional
transfer method, the initial state of the system $I_{1z}$ evolves
through the following stages
\begin{equation}
\label{eq:convinept} I_{1z} \rightarrow 2I_{1z}I_{2z} \rightarrow
2I_{2z}I_{3z} \rightarrow I_{3z}.
\end{equation}
In the first stage of the transfer, $I_{1z} \rightarrow
2I_{1z}I_{2z}$, spin $3$ is decoupled from the chain using
standard decoupling methods \cite{Ernst} and the initial
polarization $I_{1z}$ on spin $1$ is rotated by RF field (an
appropriate $\pi/2$ pulse) to coherence $I_{1x}$, which then
evolves under coupling Hamiltonian $2I_{1z}I_{2z}$ to
$2I_{1y}I_{2z}$. When the expectation value $\langle 2I_{1y}I_{2z}
\rangle $ is maximized, another $\pi/2$ pulse is used to rotate
$2I_{1y}I_{2z}$ to $2I_{1z}I_{2z}$. This is the INEPT pulse
sequence. The next stage, $2I_{1z}I_{2z}\rightarrow2I_{2z}I_{3z}$,
is the so-called spin order transfer. Fig. \ref{fig:zzlevels}
shows the population inversion corresponding to this transfer. By
a suitable $\pi/2$ rotation of spin 2, the density operator
$2I_{1z}I_{2z}$ is transformed to $2I_{1z}I_{2x}$, which then
evolves to $2I_{2x}I_{3z}$. When the expectation value $\langle
2I_{2x}I_{3z} \rangle$ is maximized, another $\pi/2$ pulse is used
to rotate $2I_{2x}I_{3z}$ to the spin order $2I_{2z}I_{3z}$. This
is the Concatenated INEPT (CINEPT) pulse sequence. The final stage
transfer, $2I_{2z}I_{3z} \rightarrow I_{3z}$, is similar to that
in the first stage and is accomplished by the INEPT pulse
sequence. The efficiency of these transfers is limited by the
decay of transverse operators, due to the phenomenon of
relaxation.

In our recent work on relaxation optimized control of coupled spin
dynamics \cite{Khaneja1}, we showed that the efficiency of the
first and the last step (the two INEPT stages) in Eq.
(\ref{eq:convinept}) can be significantly improved by controlling
precisely the way in which magnetization is transferred from
longitudinal operators to transverse operators. In other words,
instead of using $\pi/2$ (hard) pulses to rotate longitudinal to
transverse operators (and the inverse), we can exploit the fact
that longitudinal operators are long lived by making these
rotations gradually, saving this way magnetization. This transfer
strategy is called relaxation optimized pulse element (ROPE).

In this article we derive relaxation optimized pulse sequences for
the intermediate transfer (the CINEPT stage),
\begin{equation}\label{transfer2}
 2I_{1z}I_{2z} \rightarrow 2I_{2z}I_{3z}\;.
\end{equation}
This relaxation optimized transfer of spin order can then be used
as a building block for the polarization transfer (\ref{transfer})
through the scheme
\begin{equation}
\label{eq:fulltransfer} I_{1z}\rightarrow 2I_{1z}I_{2z}\rightarrow
2I_{2z}I_{3z}\rightarrow\ldots\rightarrow
2I_{(n-1)z}I_{nz}\rightarrow I_{nz}\;.
\end{equation}

\section{\label{formulation}The Optimal Control Problem and an Upper Bound for the Efficiency}

In this section, we formulate the problem of transfer in Eq.
(\ref{transfer2}) as a problem of optimal control and derive an
upper bound on the transfer efficiency. To simplify notation, we
introduce the following symbols for the expectation values of
operators that play a part in the transfer. Let $z_1 = \langle
2I_{1z}I_{2z} \rangle $, $x_1 = \langle 2I_{1z}I_{2x} \rangle$,
$y_2=\langle\sqrt{2}(2I_{1z}I_{2y}I_{3z}+I_{2y}/2)\rangle $, $x_3
= -\langle 2I_{2x}I_{3z} \rangle $ and $z_3 = \langle
2I_{2z}I_{3z} \rangle $. As a control variable we use the
transverse RF magnetic field, pointing say in the $y$ direction
(in the rotating frame), so $H_{rf}=\omega_y(t)I_{2y}$. Note that
$\omega_y(t)$ is the component of the field in the rotating frame,
so it is actually the envelope of the RF field. The carrier
frequency of the RF field is the resonance frequency of spin 2.
Using Eq. (\ref{eq:master}), we find that the evolution of the
system, in time units of $1/(\pi J\sqrt{2})$, is given by
\begin{equation}
\label{eq:path} \left[\begin{array}{ccccc}
\dot{z}_1\\\dot{x}_1\\\dot{y}_2\\\dot{x}_3\\\dot{z}_3\end{array}\right]
=\left[\begin{array}{ccccc} 0 & -\Omega_y & 0 & 0 & 0 \\
\Omega_y & -\xi & -1 & 0 & 0 \\
0 & 1 & -\xi & -1 & 0 \\
0 & 0 & 1 & -\xi & -\Omega_y \\
0 & 0 & 0 & \Omega_y & 0 \\
\end{array}\right]
\left[\begin{array}{ccccc}
z_1\\x_1\\y_2\\x_3\\z_3\end{array}\right]\;,
\end{equation}
where $\Omega_y(t)=\omega_y(t)/(\pi J\sqrt{2})$ and
$\xi=k/(J\sqrt{2})$. The initial condition is $(z_1, x_1, y_2,
x_3, z_3)=(1, 0, 0, 0, 0 )$.

The efficiency of the conventional method (CINEPT) for transfer
$z_1\rightarrow z_3$, can be easily found. At $t=0$, $z_1$ is
transferred to $x_1$ by application of a $(\pi/2)_y$ pulse on spin
2. Couplings evolve $x_1$ to $y_2$ which further evolves to $x_3$.
As a function of time, $x_3(t)=e^{-\xi t}\sin^2(t/\sqrt{2})$. This
is maximized for $t_m=\sqrt{2}\cot^{-1}(\xi/\sqrt{2})$. At $t=t_m$
a second $(\pi/2)_y$ pulse is applied on spin 2, rotating the
maximum value from $x_3$ to $z_3$. This value is the efficiency
$\eta_{CI}$ of the conventional method
\begin{equation}
\label{eq:INEPT}
\eta_{CI}=\exp(-\xi\sqrt{2}\cot^{-1}(\xi/\sqrt{2}))\sin^2(\cot^{-1}(\xi/\sqrt{2}))\;.
\end{equation}

A better efficiency can be achieved if we store magnetization in
the decoherence free longitudinal operators $z_i$ while the system
is evolving. This is done by rotating $z_i$ to $x_i$ gradually,
instead of using $\pi/2$ hard pulses. This is the physical concept
behind the relaxation optimized transfer strategy. For the
specific transfer examined in this article, we first find an upper
bound for the maximum efficiency and in the next section we
calculate numerically the magnetic field $\Omega_y(t)$ that
approaches this bound.

In order to derive the upper bound, we use an augmented system
instead of the original one (\ref{eq:path}). The augmentation is
done in two steps. First, we suppose that we can rotate $z_1$ to
$x_1$ and $x_3$ to $z_3$ independently using two different
controls, say $\Omega_1(t)$ and $\Omega_3(t)$, instead of the
common control $\Omega_y(t)$. Next, we provide $y_2$ with a
relaxation free partner $z_2$ and with a control $\Omega_2(t)$
which can rotate $y_2$ to $z_2$. The augmented system is
\begin{equation}
\label{eq:augmented} \left[\begin{array}{cccccc}
\dot{z}_1\\\dot{z}_2\\\dot{z}_3\\\dot{x}_1\\\dot{y}_2\\\dot{x}_3\end{array}\right]
=\left[\begin{array}{cccccc}
0 & 0 & 0 & -\Omega_1 & 0 & 0 \\
0 & 0 & 0 & 0 & -\Omega_2 & 0 \\
0 & 0 & 0 & 0 & 0 & -\Omega_3 \\
\Omega_1 & 0 & 0 & -\xi & -1 & 0 \\
0 & \Omega_2 & 0 & 1 & -\xi & -1 \\
0 & 0 & \Omega_3 & 0 & 1 & -\xi
\end{array}\right]
\left[\begin{array}{cccccc}
z_1\\z_2\\z_3\\x_1\\y_2\\x_3\end{array}\right]\;.
\end{equation}
Observe that system (\ref{eq:augmented}) reduces to system
(\ref{eq:path}) for $\Omega_1=-\Omega_3=\Omega_y$ and
$\Omega_2=0$. Thus, if we know the maximum achievable value of
$z_3$ starting from $(1,0,0,0,0,0)$ and evolving under system
(\ref{eq:augmented}), then this is an upper bound for the maximum
achievable value of $z_3$ with evolution described by the original
system (\ref{eq:path}).

\begin{figure}[t]
\centering
\includegraphics[scale=0.5,clip=true,bb=0.65cm 2cm 17cm 8.5cm]{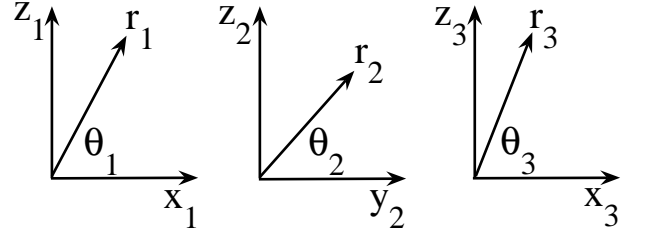}
\caption{\label{fig:angles}Auxiliary variables $r_i$.}
\end{figure}

Let $r_1 =\sqrt{x_1^2+z_1^2}$, $r_2 = \sqrt{y_2^2+z_2^2}$ and $r_3
= \sqrt{x_3^2+z_3^2}$. Using $\Omega_i(t)$ we can control the
angles $\theta_i$, shown in Fig. \ref{fig:angles}, independently.
If we assume that the control can be done arbitrarily fast as
compared to the evolution of couplings or relaxation rates then we
can think of $\theta_i$ as control variables. The equations for
the evolution of $r_i$ are
\begin{equation}
\label{eq:systemr} \left[\begin{array}{ccc}
\dot{r}_1\\\dot{r}_2\\\dot{r}_3\end{array}\right]
= \left[\begin{array}{ccc}-\xi u_1^2 & -u_1u_2 & 0 \\
u_1u_2 & -\xi u_2^2 & -u_2u_3 \\
0 & u_2u_3 & -\xi u_3^2\end{array}\right] \left[\begin{array}{ccc}
r_1\\r_2\\r_3\end{array}\right],
\end{equation}
where the new control parameters are $u_i(t) = \cos(\theta_i(t))$.
The goal is to find the largest achievable value of $r_3$ starting
from $(r_1, r_2, r_3) = (1, 0, 0)$ by appropriate choice of
$u_i(t)$. This problem can be solved analytically. The optimal
solution is characterized by maintaining vanishingly small values
of $dr_2/dt$, i.e., $(u_1r_1 - u_3r_3)/u_2r_2= \xi$ and by
$u_3r_3/u_1r_1 = \kappa$, where
\begin{equation}\label{eq:upperbound}
\kappa = \frac{(\sqrt{\xi^2 + 2}-\xi)^2}{2}.
\end{equation}
The maximum achievable value of $r_3$ is also $\kappa$. We prove
it in the following.

Using variables $p_i = {r_i^2}$, $m_i = u_ir_i/\sqrt{\sum_{i}^3
(u_ir_i)^2}$, and $d\tau/dt = \sum_{i}^3 (u_ir_i)^2$ , equation
(\ref{eq:systemr}) can be re-written as
\begin{equation}\label{eq:psystem}
\frac{d}{d \tau} \left[\begin{array}{c} {p}_1(\tau)\\ {p}_2(\tau)
\\ {p}_3(\tau) \end{array}\right] = \ diag (A m(\tau)m^T(\tau))\;,
\end{equation}
where
\begin{equation}\label{eq:A}
A = 2 \left[\begin{array}{ccc}-\xi & -1 & 0 \\
1 & -\xi & -1 \\
0 & 1 & -\xi \end{array}\right],
\end{equation} $m^T = (m_1, m_2, m_3)$ and $diag(X)$ represents the vector containing
diagonal entries of the square matrix $X$. The goal is to find the
controls $m_i(\tau)$ ($\sum_i m_i^2(\tau) = 1$) and the largest
achievable value of $p_3$ starting from $(p_1, p_2, p_3) = (1, 0,
0)$. Eq. (\ref{eq:psystem}) implies that
$$\left[\begin{array}{c}
{p}_1(T)\\ {p}_2(T) \\ {p}_3(T) \end{array}\right] =
\left[\begin{array}{c} {p}_1(0)\\ {p}_2(0) \\ {p}_3(0)
\end{array}\right] + diag (A \int_0^T m(\tau)m^T(\tau)d \tau)\;. $$
Let $M =  \int_0^T m(\tau)m^T(\tau) d\tau$. Note that $M$ is a
symmetric, positive semidefinite matrix. By definition $p_i(\tau)
\geq 0$ for all $\tau$. At final time $T$, we must have $p_1(T)=0$
and $p_2(T)=0$, as any nonzero value of $p_1(T)$ or $p_2(T)$ can
be partly transferred to $p_3$ and the final value of $p_3(T)$
further increased. Since $(p_1(0), p_2(0), p_3(0)) = (1, 0, 0)$,
it implies that $M$ should be such that $(AM)_{11} = -1$,
$(AM)_{22} = 0$ and $(AM)_{33}$ is maximized over all positive
semidefinite $M$ satisfying the above constraints. This problem is
a special case of a semidefinite programming problem \cite{Boyd}.
If the symmetric part of matrix $A$ is negative definite, as in
our case (\ref{eq:A}), then it can be shown that the optimal
solution $M$ to the above stated semi-definite programming problem
is a rank one matrix \cite{dionisis}, i.e., $M = mm^{T}$ for some
constant $m$ and therefore the ratio $u_ir_i/u_jr_j$ in
(\ref{eq:systemr}) is constant throughout. The condition
$(AM)_{22} = 0$, implies $(u_1r_1 - u_3r_3)/\xi = u_2r_2$.
Substituting for $u_2r_2$ in (\ref{eq:systemr}), we obtain
 \begin{equation}
\label{eq:systemr.1} \left[\begin{array}{cc}
\dot{r}_1\\\dot{r}_3\end{array}\right]
= \left[\begin{array}{cc}-(\xi + 1/\xi) u_1^2 & u_1u_3/\xi \\
u_1u_3/\xi & -(\xi + 1/\xi) u_3^2
\end{array}\right] \left[\begin{array}{ccc}
r_1\\r_3\end{array}\right],
\end{equation}

We now simply need to maximize the gain in $r_3$ to loss in $r_1$,
i.e., the ratio  $\dot{r}_3/(-\dot{r}_1)$. This yields
$u_3r_3/u_1r_1= \kappa$, with $\kappa$ given in
(\ref{eq:upperbound}). The corresponding maximum efficiency for
transfer $r_1\rightarrow r_3$ is also $\kappa$. This is the
maximum efficiency for transfer $z_1\rightarrow z_3$ under the
augmented system (\ref{eq:augmented}), and thus an upper bound for
the efficiency of the same transfer under the original system
(\ref{eq:path}).

\begin{figure*}[t]
\centering
\begin{tabular}{cc}
(a) \psfig{file=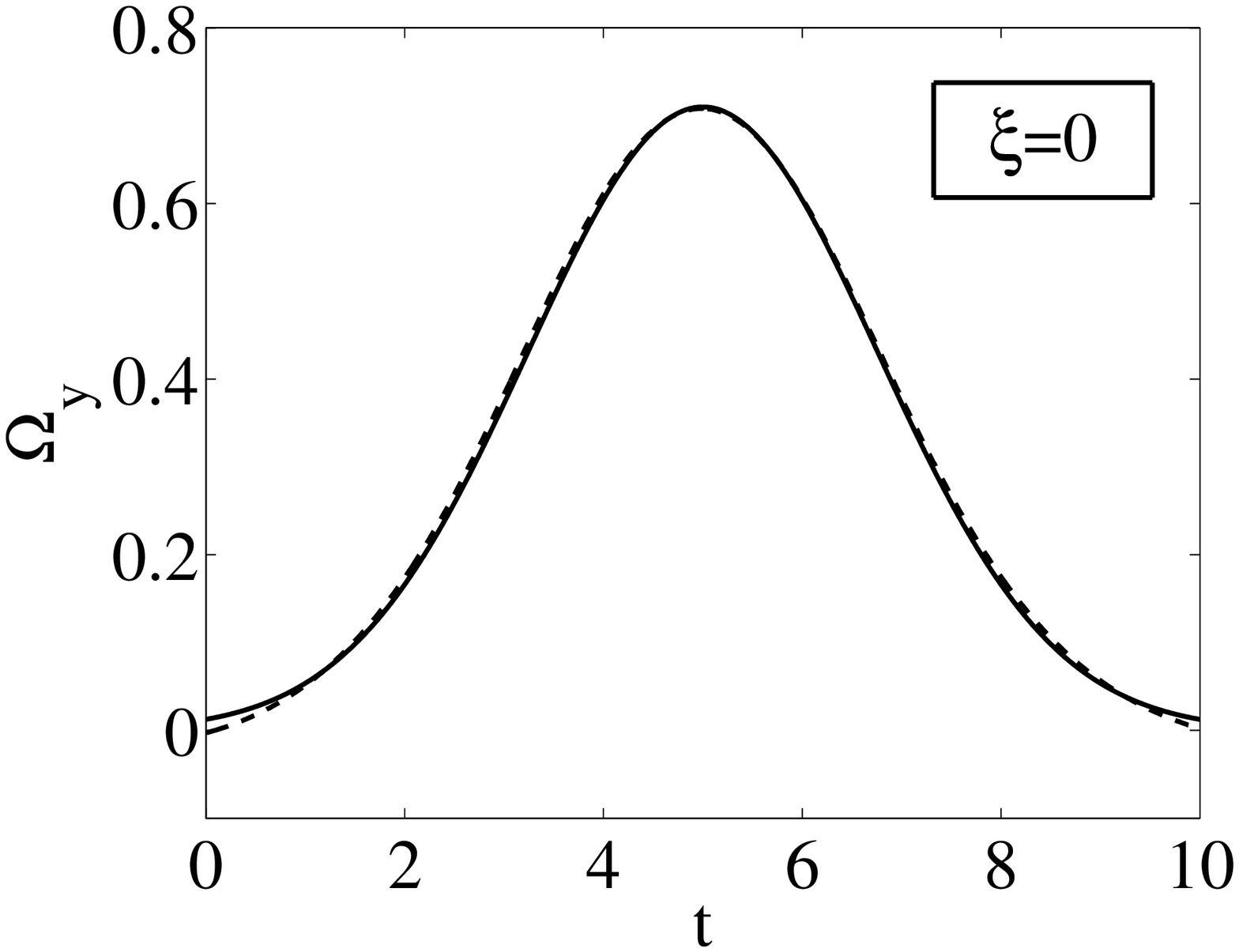,scale=0.4} &
(b) \psfig{file=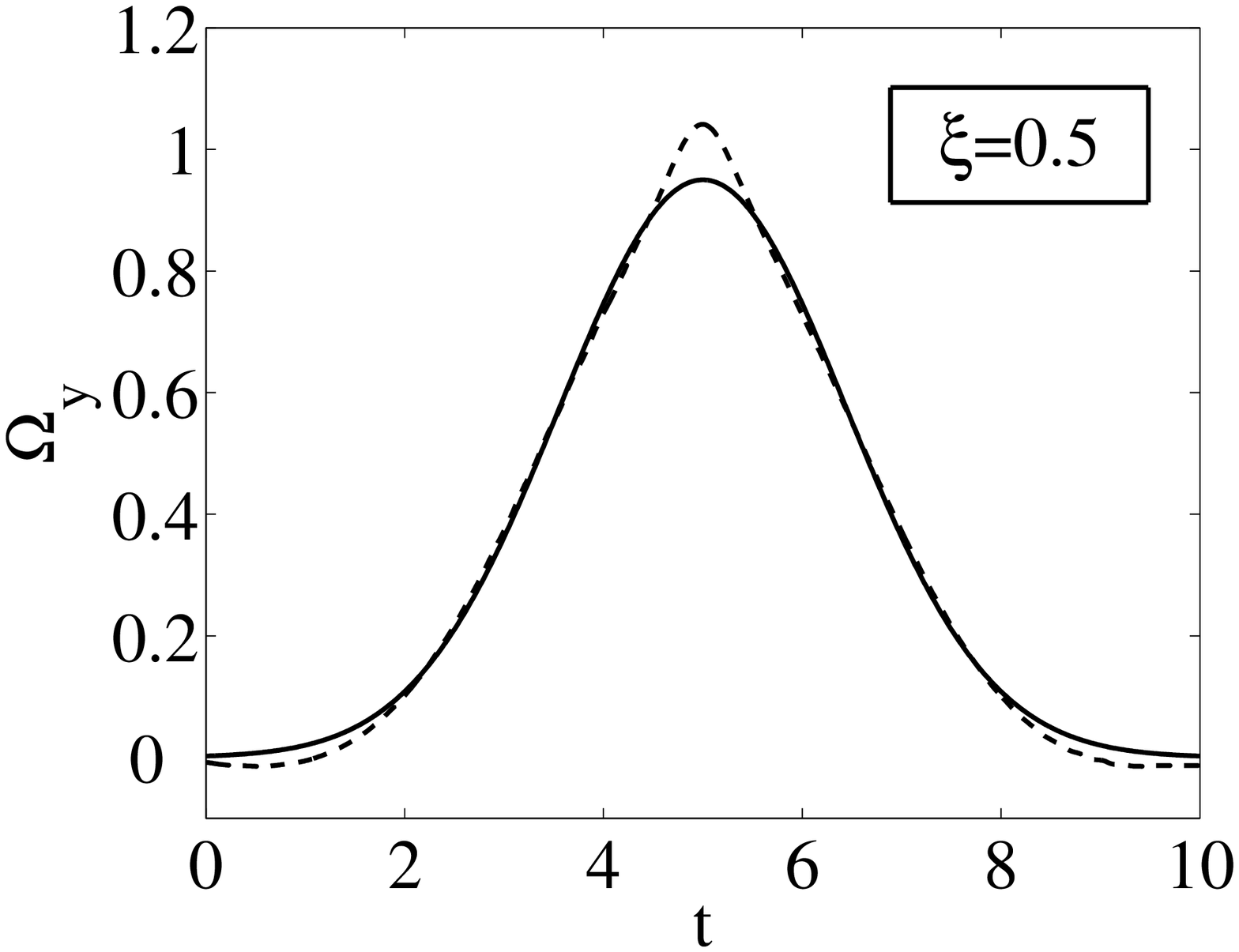,scale=0.4} \\
(c) \psfig{file=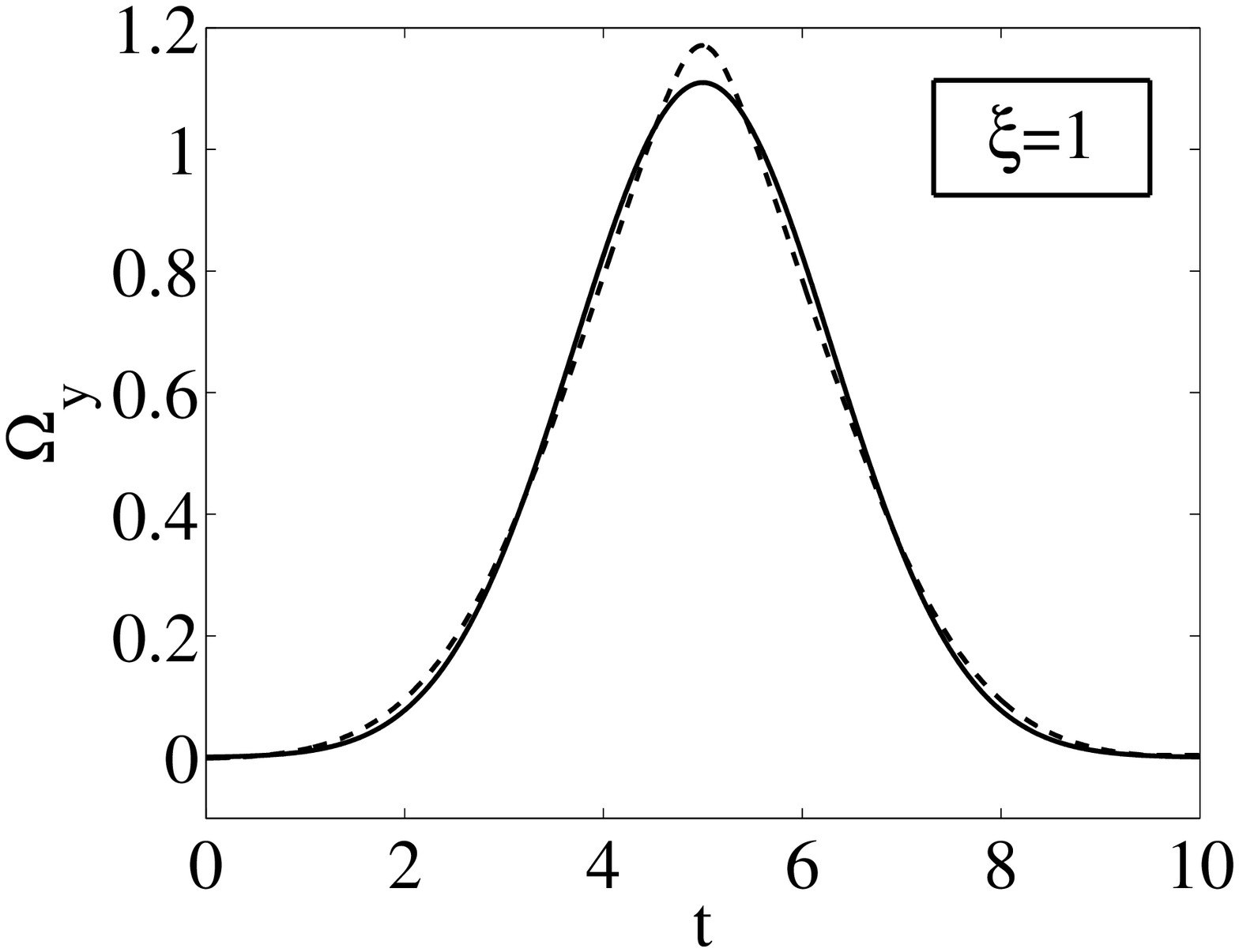,scale=0.4} &
(d) \psfig{file=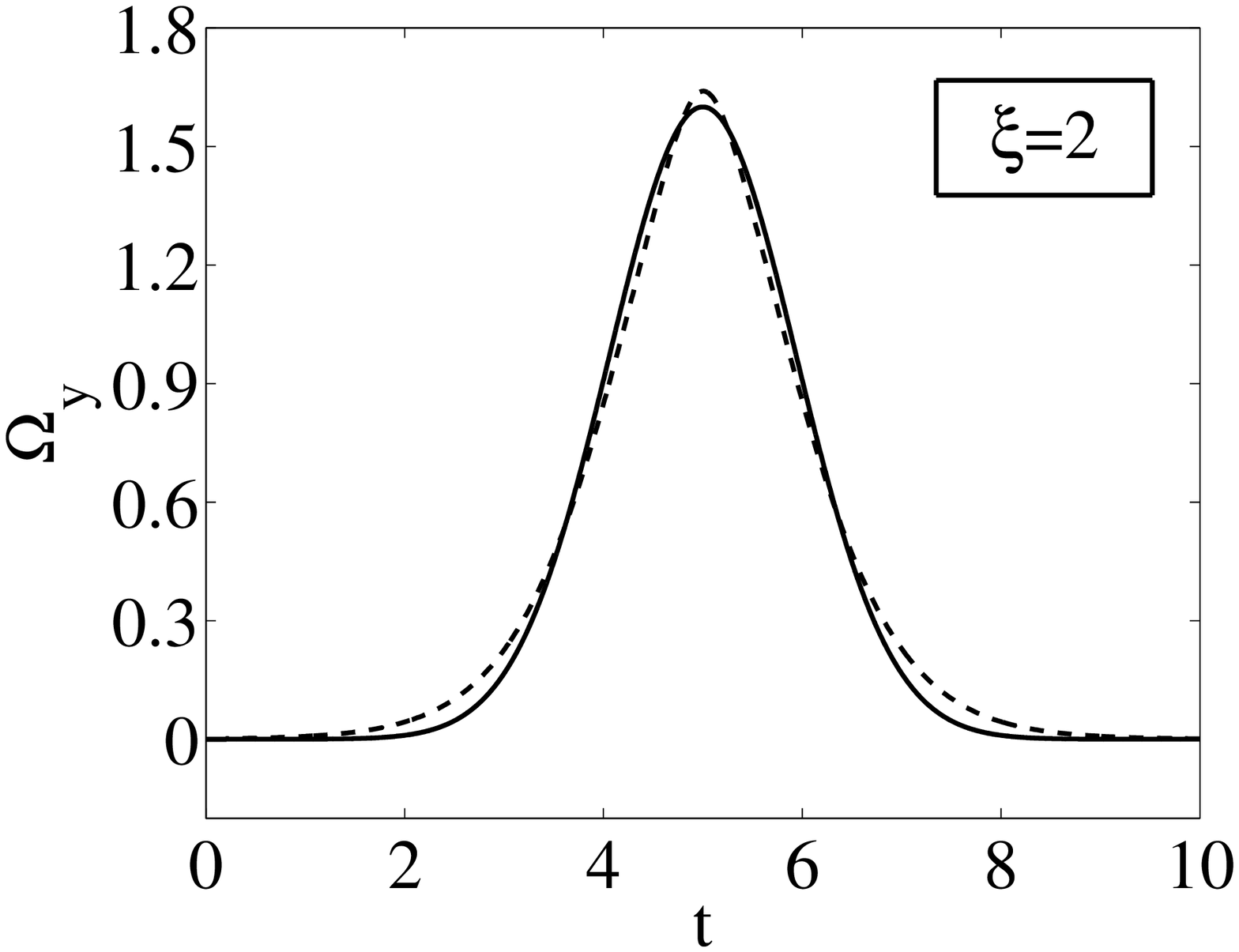,scale=0.4} \\
(e) \psfig{file=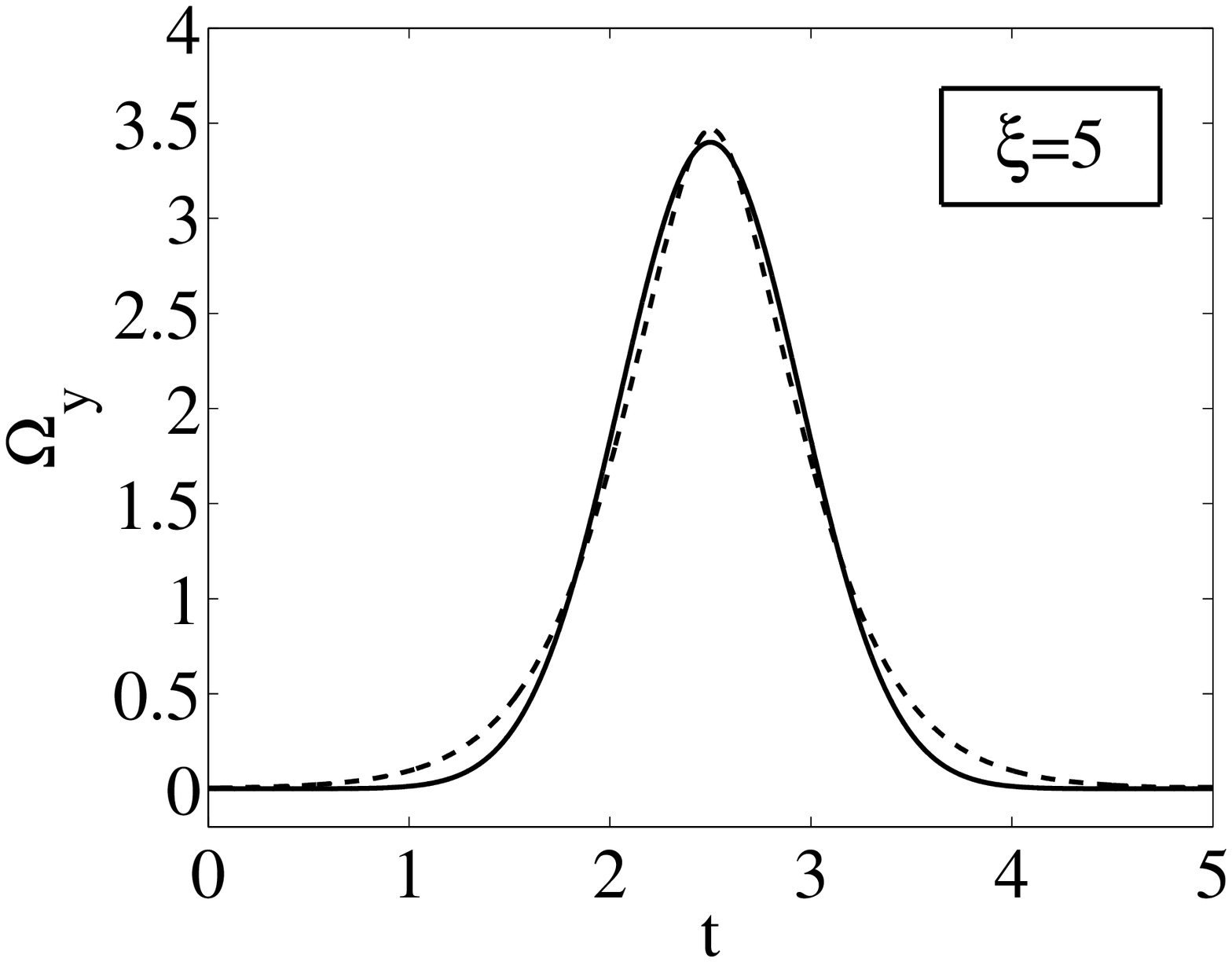,scale=0.4} & \quad (f)\quad
\psfig{file=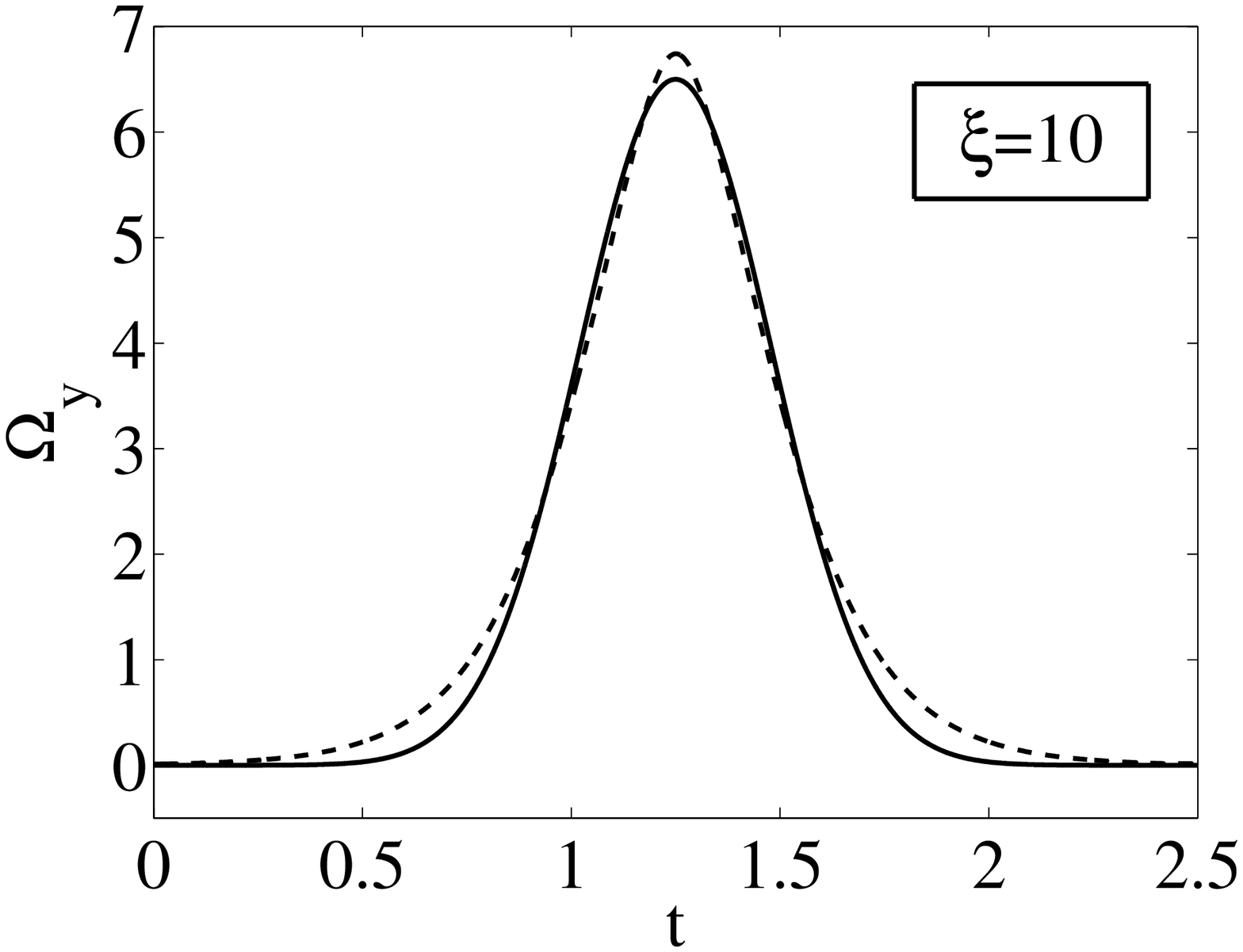,scale=0.4}
\end{tabular}
\caption{\label{fig:optimalfield}Optimal pulse (dashed line)
calculated using a numerical optimization method based on a
steepest descent algorithm, for various values of the normalized
relaxation parameter $\xi$.  The Gaussian pulse (solid line)
approximates very well the optimal pulse shape and gives a similar
efficiency. This suggests that instead of using the initial
numerical optimization method, we can use Gaussian pulses of the
form (\ref{gaussian}), optimized with respect to $A$ and $\sigma$
for each value of $\xi$.}
\end{figure*}

\section{\label{results} Numerical Calculation of the Optimal RF field and Discussion}

Having established an analytical upper bound (\ref{eq:upperbound})
for the efficiency, we now try to find numerically a RF field
$\Omega_y(t)$ that approaches this bound, for each value of the
parameter $\xi$. We emphasize that in this section the original
system (\ref{eq:path}) is employed.

At first, we use a numerical optimization method based on a
steepest descent algorithm. For the application of the method we
use a finite time window $T$. For values of normalized relaxation
$\xi$ in $(0-1)$, a time interval $T=10$ (normalized time units)
is enough. For larger values of $\xi$ we can use even shorter $T$.
The optimal RF field $\Omega_y(t)$ that we find with this method,
for various $\xi$, is shown in Fig. \ref{fig:optimalfield}. Note
that as $\xi$ increases, the optimal pulse becomes shorter in time
and acquires a larger peak value. The reason for this is that for
larger $\xi$ the transfer $z_1\rightarrow z_3$ should be done
faster, in order to reduce the time spent in the transverse plane
and hence the relaxation losses. Now observe that the optimal
pulse shape can be very well approximated by a Gaussian profile of
the form
\begin{equation}
\label{gaussian}
\Omega_y(t)=A\exp{\left[\left(\frac{t-T/2}{\sqrt{2}\sigma}\right)^2\right]}\;,
\end{equation}
with $A, \sigma$ appropriately chosen. As a result, the efficiency
that we find using the appropriate Gaussian pulse is very close to
that we find using the original pulse. This suggests that instead
of using the initial numerical optimization method, we can use
Gaussian pulses of the form (\ref{gaussian}), optimized with
respect to $A$ and $\sigma$ for each value of $\xi$. The optimal
$A, \sigma$ are found by numerical simulations. For each $\xi$ we
simulate the equations of system (\ref{eq:path}) with
$\Omega_y(t)$ given by Eq. (\ref{gaussian}), for many values of
$A$ and $\sigma$. We choose those values that give the maximum
$z_3(T)$. In table I, we show the optimal $A, \sigma$ for various
values $\xi\in [0, 1]$. We also show the corresponding efficiency,
as well as the efficiency achieved by the initial numerical
optimization method. Observe how close lie these two groups of
values. The choice of the Gaussian shape is indeed successful.
\begin{table}
\caption{For various values of $\xi\in[0,1]$, the optimal values
of $A,\sigma$ and the corresponding efficiency are shown. We
present also for comparison the efficiency achieved by the
steepest descent method.}
\begin{ruledtabular}
\begin{tabular}{ccccc}
$\xi$ &$A$ &$\sigma$
& $\mbox{Gaussian Pulse}$ & $\mbox{Steepest Descent}$\\
\hline
1.00 & 1.11 & 1.30 & 0.2510 & 0.2512\\
0.95 & 1.09 & 1.32 & 0.2661 & 0.2662\\
0.90 & 1.07 & 1.34 & 0.2824 & 0.2825\\
0.85 & 1.05 & 1.36 & 0.3000 & 0.3001\\
0.80 & 1.03 & 1.38 & 0.3190 & 0.3191\\
0.75 & 1.02 & 1.39 & 0.3396 & 0.3397\\
0.70 & 1.00 & 1.41 & 0.3619 & 0.3620\\
0.65 & 0.98 & 1.43 & 0.3861 & 0.3863\\
0.60 & 0.97 & 1.44 & 0.4124 & 0.4126\\
0.55 & 0.96 & 1.44 & 0.4410 & 0.4413\\
0.50 & 0.95 & 1.44 & 0.4721 & 0.4726\\
0.45 & 0.94 & 1.45 & 0.5060 & 0.5067\\
0.40 & 0.93 & 1.46 & 0.5428 & 0.5439\\
0.35 & 0.92 & 1.46 & 0.5830 & 0.5846\\
0.30 & 0.91 & 1.46 & 0.6270 & 0.6292\\
0.25 & 0.90 & 1.47 & 0.6750 & 0.6780\\
0.20 & 0.89 & 1.48 & 0.7277 & 0.7315\\
0.15 & 0.88 & 1.48 & 0.7855 & 0.7900\\
0.10 & 0.85 & 1.52 & 0.8494 & 0.8536\\
0.05 & 0.79 & 1.60 & 0.9203 & 0.9232\\
0.00 & 0.73 & 1.71 & 0.9999 & 1.0000
\end{tabular}
\end{ruledtabular}
\end{table}

\begin{figure}[h]
\centering
\includegraphics[scale=0.4]{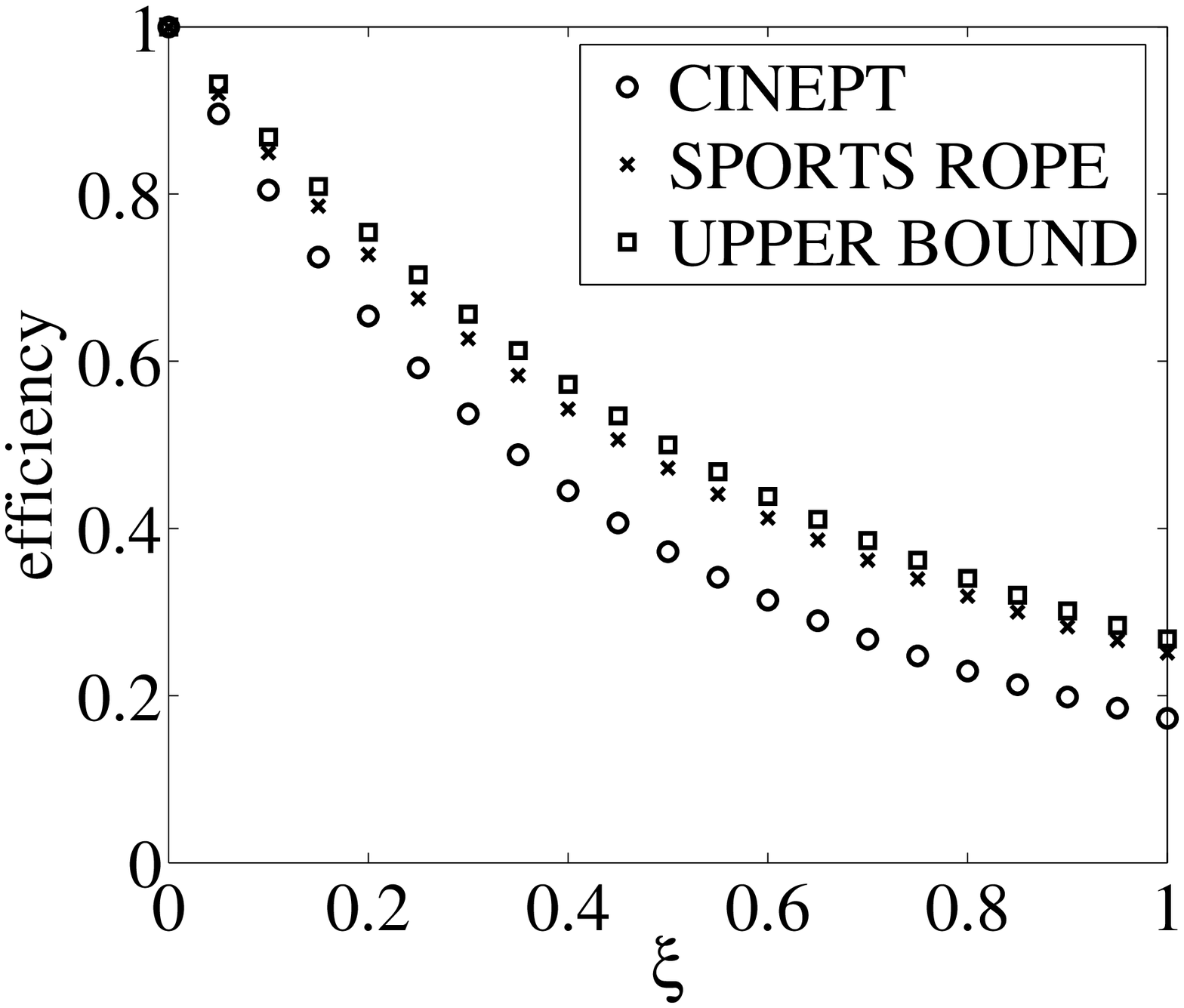}
\caption{\label{fig:efficiency}Efficiency for the conventional
method (CINEPT), Eq. (\ref{eq:INEPT}), and for our method (SPORTS
ROPE), for the values of $\xi$ shown in table I. The upper bound
(\ref{eq:upperbound}) for the efficiency is also shown.}
\end{figure}

Fig. \ref{fig:efficiency} shows the efficiency of the conventional
method (CINEPT), i.e., $\eta_{CI}$ from Eq. (\ref{eq:INEPT}), the
efficiency of our method (SPORTS ROPE, SPin ORder TranSfer with
Relaxation Optimized Pulse Element), and the upper bound $\kappa$
from Eq. (\ref{eq:upperbound}), for the values of relaxation
parameter $\xi$ shown in table I. Note that for large $\xi$ (large
relaxation rates), SPORTS ROPE gives a significant improvement
over CINEPT. Also note that it approaches fairly well the upper
bound.

\begin{figure}[h]
\centering
\includegraphics[scale=0.4]{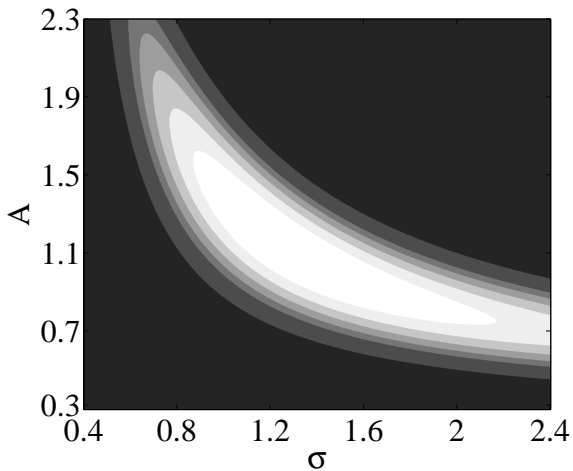}
\caption{\label{fig:robustness}Gray-scale topographic plot of the
efficiency as a function of the amplitude $A$ and the standard
deviation $\sigma$ of the Gaussian pulse for $\xi=1$. The maximum
value can be found from table I and is 0.2510. The white region
corresponds to values $\geq 0.24$, while the black region to
values $<\eta_{CI}$($\xi$=1) = 0.1727. The intermediate gray
regions correspond to the intervals
$[0.23,0.24),[0.22,0.23),[0.21,0.22),[0.20,0.21),[0.1727,0.20)$
(from white to black).}
\end{figure}

Using the Gaussian pulse shape we can get a quantitative
impression of the robustness of SPORTS ROPE. In Fig.
\ref{fig:robustness} we give a gray-scale topographic plot of the
efficiency ($z_3(T)$ with $T=10$) as a function of $A$ and
$\sigma$ for $\xi=1$. The maximum value can be found from table I
and is 0.2510. The white region corresponds to values $\geq 0.24$,
while the black region to values $<\eta_{CI}(\xi=1)=0.1727$. The
intermediate gray regions correspond to values between these two
limits. Obviously, SPORTS ROPE is quite robust.

\begin{figure*}
\centering
\begin{tabular}{cc}
(a) \psfig{file=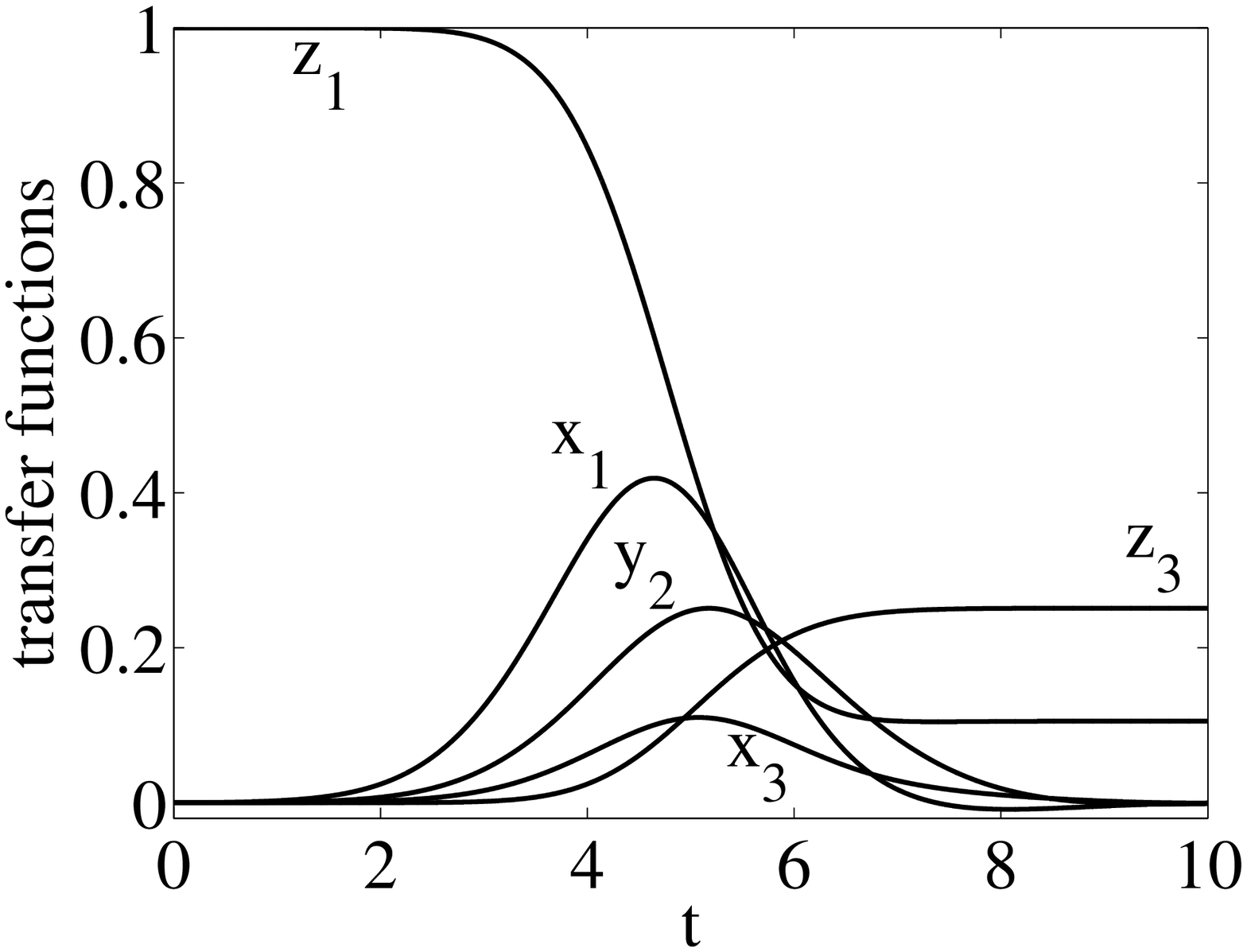,scale=0.4} & (b)
\psfig{file=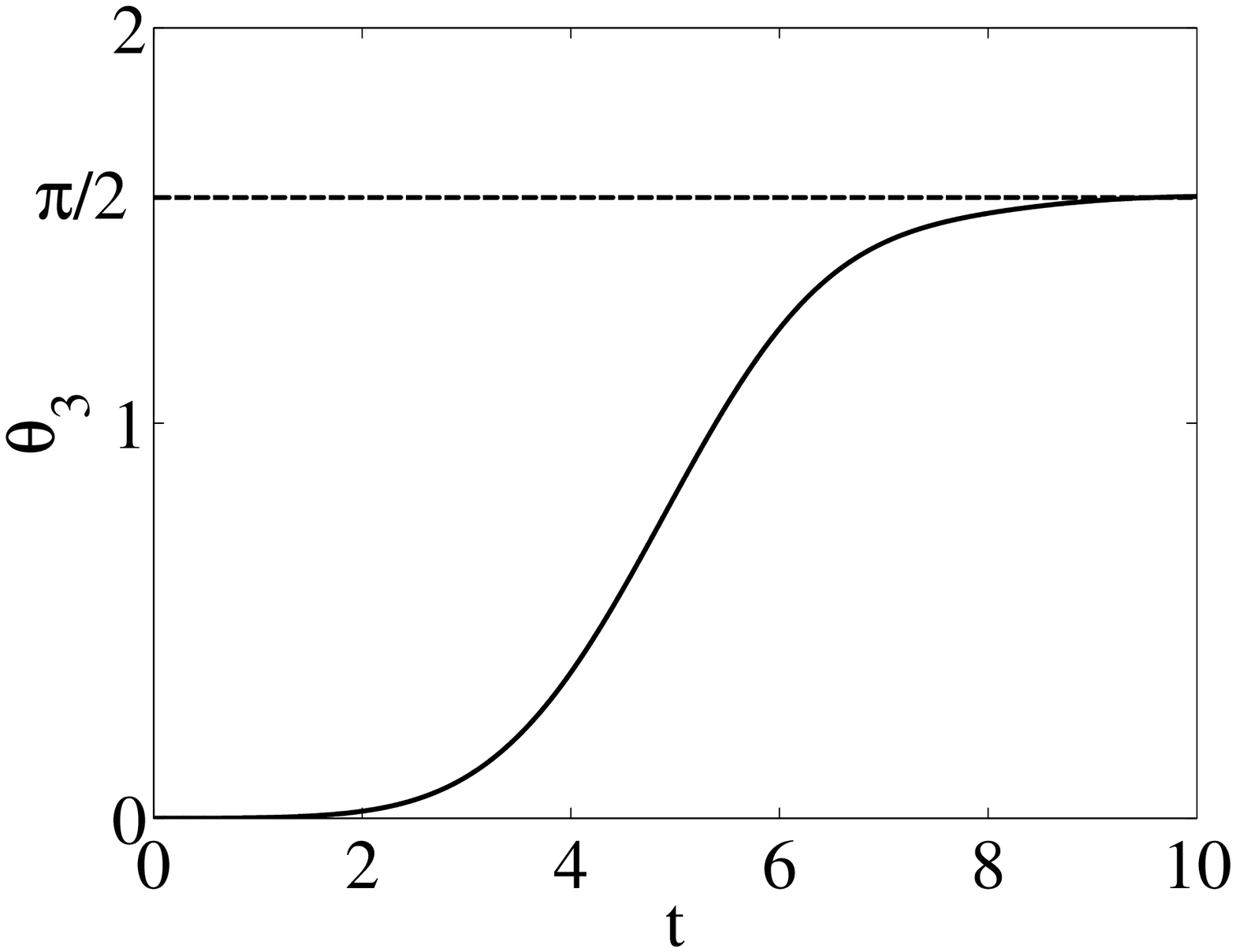,scale=0.4}
\end{tabular}
\caption{\label{fig:populations}(a) Time evolution of the various
transfer functions (expectation values of operators) participating
in the transfer $z_1\rightarrow z_3$, when the optimal Gaussian
pulse for $\xi=1$, shown in Fig. \ref{fig:optimalfield}(c), is
applied to system (\ref{eq:path}). (b) The angle
$\theta_3=\tan^{-1}{(z_3/x_3)}$ of the vector
$\textit{\textbf{r}}_3$ with the $x$ axis, as a function of time.
Observe that initially $\textit{\textbf{r}}_3$ is parallel to $x$
axis ($\theta_3=0$), but under the action of the Gaussian pulse is
rotated gradually to $z$ axis ($\theta_3=\pi/2$).}
\end{figure*}

In Fig. \ref{fig:populations}(a), we plot the time evolution of
the various transfer functions (expectation values of operators)
that participate in the transfer $z_1\rightarrow z_3$, when the
optimal Gaussian pulse for $\xi=1$, shown in Fig.
\ref{fig:optimalfield}(c), is applied to system (\ref{eq:path}).
Observe the gradual building of the intermediate variables $x_1,
y_2$ and $x_3$. Note that $dr_2/dt=\dot{y}_2\neq 0$. There is no
contradiction with the optimality condition $dr_2/dt=0$ derived in
section \ref{formulation} during the calculation of the upper
bound, since this condition refers to the augmented system
(\ref{eq:augmented}) and not the original one (\ref{eq:path}) used
here. In Fig. \ref{fig:populations}(b) we plot the angle
$\theta_3=\tan^{-1}{(z_3/x_3)}$ of the vector
$\textit{\textbf{r}}_3$ with the $x$ axis, as a function of time.
Observe that initially $\textit{\textbf{r}}_3$ is parallel to $x$
axis ($\theta_3=0$), but under the action of the Gaussian pulse is
rotated gradually to $z$ axis ($\theta_3=\pi/2$). This gradual
rotation of $\textit{\textbf{r}}_3$ (as well as of
$\textit{\textbf{r}}_1$) is a characteristic feature of the SPORTS
ROPE transfer scheme.

We remark that for the general transfer $I_{1z}\rightarrow
I_{nz}$, more than one intermediate steps
$2I_{(i-1)z}I_{iz}\rightarrow 2I_{iz}I_{(i+1)z}$ are necessary.
Since the equations that describe the $i^{th}$ transfer are the
same as $(\ref{eq:path})$, we just need to apply the same Gaussian
pulse but centered, in the frequency domain, at the resonance
frequency of spin $i$. In this sequence of Gaussian pulses we
should add at the beginning and at the end the optimal pulses for
the first and the final step, respectively, see Eq.
(\ref{eq:fulltransfer}). These pulses can be calculated using the
theory presented in \cite{Khaneja1}. Finally, note that the same
scheme can be used for the coherence transfer
$I_{1\alpha}\rightarrow I_{n\beta}$, where $\alpha,\beta$ can be
$x$ or $y$. We just need to add the initial and final $\pi/2$
pulses that accomplish the rotations $I_{1\alpha}\rightarrow
I_{1z}$, $I_{nz}\rightarrow I_{n\beta}$.

\section{\label{conc}Conclusion}

In this paper, we derived an upper bound on the efficiency of spin
order transfer along an Ising spin chain, in the presence of
relaxation, and calculated numerically relaxation optimized pulse
sequences approaching this bound. Using these methods, a
significant reduction in relaxation losses is achieved, compared
to standard techniques, when transverse relaxation rates are much
larger than the longitudinal relaxation rates and comparable to
couplings between spins. These relaxation optimized methods can be
used as a building block for transfer of polarization or coherence
between distant spins on a spin chain. This problem is ubiquitous
in multi-dimensional NMR spectroscopy and is also interesting in
the context of quantum information processing.

\begin{acknowledgments}
N.K. acknowledges Grants AFOSR FA9550-04-1-0427, NSF 0133673 and
NSF 0218411. S.J.G. thanks the Deutsche Forschungsgemeinschaft for
Grant Gl 203/4-2.
\end{acknowledgments}


\begin{thebibliography}{99}

\bibitem{Khaneja1} N. Khaneja, T. Reiss, B. Luy, and S.J. Glaser, J. Magn.
Reson. 162, 311 (2003).

\bibitem{Khaneja2} N. Khaneja, B. Luy, and S.J. Glaser (2003) Proc.
Natl. Acad. Sci. U.S.A 100, 13162 (2003).

\bibitem{Stefanatos} D. Stefanatos, N. Khaneja, and S.J. Glaser,
Phys. Rev. A, 69, 022319 (2004).

\bibitem{BBCROP}N. Khaneja, Jr. Shin Li, C. Kehlet, B. Luy, S.J. Glaser,
Proc. Natl. Acad. Sci. USA. 101, 14742-47 (2004).

\bibitem{kane} B.E. Kane, Nature {\bf 393}, 133 (1998).

\bibitem{yama} F. Yamaguchi, Y. Yamamoto, {\it Appl. Phys. A} {\bf 68} (1999).

\bibitem{NMR}J. Cavanagh, W.J. Fairbrother, A.G. Palmer III, and N.J.
Skelton, Protein NMR Spectroscopy (Academic Press, New York,
1996).

\bibitem{Goldman1}M. Goldman, Interference effects in the relaxation
of a pair of unlike spin-1/2 nuclei,  J. Magn. Reson. 60, 437
(1984).

\bibitem{inept1}
G. A. Morris, R. Freeman, {\it J. Am. Chem. Soc.} {\bf 101}, 760 (1979).

\bibitem{inept2}
D. P. Burum, R. R. Ernst, {\it J. Magn. Reson.} {\bf 39}, 163 (1980).

\bibitem{concat}
A. Majumdar, E. P. Zuiderweg, {\it J. Magn. Reson. A} {\bf 11}3, 19-31 (1995).

\bibitem{Ernst} R.R. Ernst, G. Bodenhausen, A. Wokaun, Principles
of Nuclear Magnetic Resonance in One and Two Dimensions, Clarendon
Press, Oxford, 1987.

%\bibitem{Alicki} R. Alicki, K. Lendi, Quantum Dynamical Semigroups
%and Applications, Lecture Notes in Physics, Springer, Berlin, Vol.
%286, (1987).

\bibitem{Boyd} L. Vandenberghe, S. Boyd, {\it SIAM Review} {\bf 38}, 49-95 (1996).

\bibitem{dionisis} D. Stefanatos and N. Khaneja, {\it math.OC/0504308}, (2005).

\end{thebibliography}
\end{document}